\documentclass[journal=nalefd,manuscript=review]{achemso}

\usepackage[version=3]{mhchem} 
\usepackage{color}

\author{Xiaochang Miao}
\affiliation[Department of Physics, University of Florida]
{Department of Physics, University of Florida, Gainesville, FL 32611}
\author{Sefaattin Tongay}
\affiliation[Department of Physics, University of Florida]
{Department of Physics, University of Florida, Gainesville, FL 32611}
\alsoaffiliation[Department of Material Science and Engineering, University of Florida]
{Department of Material Science and Engineering, University of Florida, Gainesville, FL 32611}
\email{tongay@ufl.edu}
\author{Maureen K. Petterson}
\affiliation[Department of Physics, University of Florida]
{Department of Physics, University of Florida, Gainesville, FL 32611}
\author{Kara Berke}
\affiliation[Department of Physics, University of Florida]
{Department of Physics, University of Florida, Gainesville, FL 32611}
\author{Andrew G. Rinzler}
\affiliation[Department of Physics, University of Florida]
{Department of Physics, University of Florida, Gainesville, FL 32611}
\author{Bill R. Appleton}
\affiliation[Nanoscience Institute for Medical and Engineering Technologies, University of Florida]
{Nanoscience Institute for Medical and Engineering Technologies, University of Florida, Gainesville, FL 32611}
\alsoaffiliation[Department of Material Science and Engineering, University of Florida]
{Department of Material Science and Engineering, University of Florida, Gainesville, FL 32611}
\author{Arthur F. Hebard}
\affiliation[Department of Physics, University of Florida]
{Department of Physics, University of Florida, Gainesville, FL 32611}
\email{afh@phys.ufl.edu}

\title[\texttt{achemso} demonstration]
{High Efficiency Graphene Solar Cells by Chemical Doping}

\begin{document}
\footnote{X. Miao and S. Tongay contributed to this work equally.}
\begin{abstract}
We demonstrate single layer graphene/\thinspace \textit{n}-Si Schottky junction solar cells that under AM1.5 illumination exhibit a power conversion efficiency (PCE) of 8.6\%. This performance, achieved by doping the graphene with bis(trifluoromethanesulfonyl)amide, exceeds the native (undoped) device performance by a factor of 4.5 and the best previously reported PCE in similar devices by a factor of nearly 6. Current-voltage, capacitance-voltage and external quantum efficiency measurements show the enhancement to be due to the doping induced shift in the graphene chemical potential which increases the graphene carrier density (decreasing the cell series resistance) and increases the cell's built-in potential (increasing the open circuit voltage) both of which improve the solar cell fill factor.
\end{abstract}

\section{Introduction}

Thin, transparent, electrically conducting films of carbon nanotubes deposited on $n$-type silicon have recently been shown to form reasonably efficient Schottky junction solar cells\cite{dehai}. Subsequently, the low density of electronic states which permits electronic or chemical charge transfer modulation of the nanotube chemical potential was exploited to show dramatically improved performance in such devices\cite{pooja,dehai2}. More recently, graphene based Schottky junction solar cells have been demonstrated on various semiconducting substrates such as Si\cite{Li1,cronin}, CdS~\cite{Ye}, and CdSe~\cite{Zhang} with power conversion efficiencies (PCE) ranging from 0.1\% up to 2.86\%. Here, we demonstrate single layer graphene/$n$-Si Schottky junction solar cells that exhibit a native (undoped) power conversion efficiency under one sun AM1.5G illumination of 1.9\%, which upon chemical charge transfer doping with bis(trifluoromethanesulfonyl)amide[((CF$_3$SO$_2$)$_2$NH)] (TFSA) increases the device PCE to 8.6\%. The TFSA dopant has the added advantage of environmental stability due to its hydrophobic nature.~\cite{tongaynano} Through independent measurements and simple modeling the several effects of the doping and means by which this improvement occurs are readily explained. We note that the graphene based Schottky junction solar cells are advantagous compared to indium tin oxide (ITO) Si junctions, since (a) graphene's work function, and hence the device properties, can be tuned as desired to optimize the solar cell efficiency, (b) graphene electrodes promise an inexpensive and convenient way to form Schottky junctions, (c) graphene is expected to outperform ITO electrically and optically, and (d) the presented technique can be applied on other semiconductors (GaAs, CdSe, etc.) owing to the tunability of graphene's work function.


We attribute the observed enhancements in the PCE to both: (1) an increase in the Schottky barrier height (SBH), and hence the built-in potential ($V_{bi}$) as measured by two complementary techniques, current density-voltage ($J$-$V$) and capacitance-voltage ($C$-$V$) and (2) a reduction in resistive losses associated with an increase in the electrical conductivity of the doped graphene sheets.~\cite{tongaynano} Besides the dramatic improvement in the PCE of such cells, our results also address the physics governing electrical transport across the graphene/$n$-Si interface. Such understanding is critical to further improvement of similar devices. 

\begin{figure}[t]
\includegraphics[angle=0,width=0.75\textwidth]{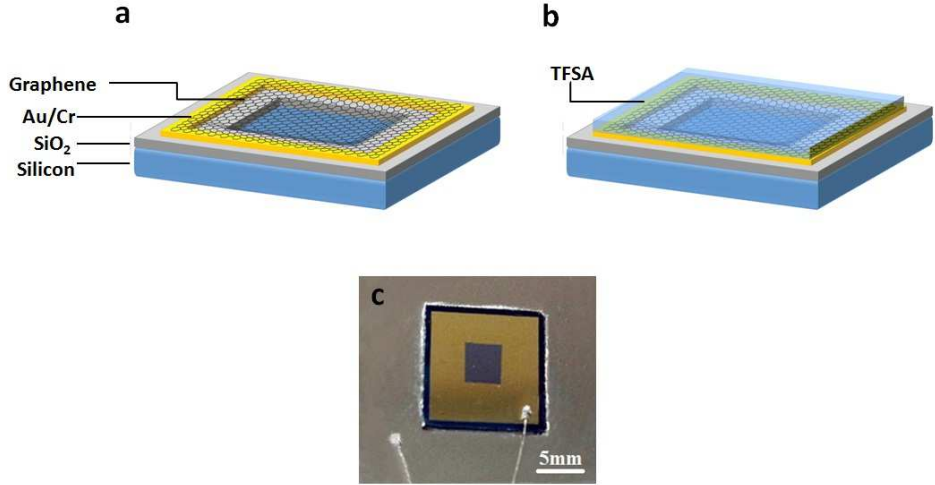}
\caption{(a) Graphene/\thinspace $n$-Si (b) TFSA doped graphene/\thinspace $n$-Si Schottky solar cell geometry (c) Optical image of a completed TFSA doped graphene/$n$-Si solar cell showing contacts and contact leads.} 
\label{Diagram}
\end{figure}



The graphene sheets used in these studies were grown on copper (Cu) foils by chemical vapor deposition (CVD)~\cite{Ruoff}. PMMA (Poly(methyl methacrylate)), used as a physical support for graphene during its removal from the Cu substrate, was spin-cast on graphene/Cu, and unwanted graphene was removed from the back side of the Cu foils by reactive ion etching (RIE)~\cite{Ruoff2}. The copper layer was removed from the PMMA/graphene/Cu foils using a Fe(NO$_3$)$_3$ etchant solution, yielding PMMA/graphene sheets. Prior to transferring graphene to a new substrate, Au/Cr windows (\ref{Diagram}a-c) were deposited onto as-received Si(111) (\textit{n}-type 8$\times$10$^{14}$-1$\times$10$^{15}$ cm$^{-3}$) wafers with a 1~$\mu$m-thick thermal oxide (SiO$_2$) surface layer. Here the gold layer provides a low resistance contact to the graphene sheets. After the Au/Cr deposition, exposed parts (3 $\times$ 3~mm area) of the SiO$_2$ were removed using a buffered oxide etch (BOE) with NH$_4$F:HF (6:1) ratio for 10 minutes to expose the underlying Si. In similar nanotube/\textit{n}-Si devices it was found that exposure of the underlying Si to ambient air for ~two hours prior to testing was beneficial for device performance.\cite{pooja2} This was likely due to oxygen passivation of dangling bonds that reduce surface states as reported for conventional MIS cells.\cite{aberle} A similar benefit was found for our graphene/\textit{n}-Si devices for which processing included up to 2 hours exposure to ambient air between etching the window and transfer of the graphene to the Si surface. Graphene sheets were transferred onto Si (\ref{Diagram}), and the PMMA backing layer was dissolved away in an acetone vapor bath and subsequently soaked for 12 hours in an acetone solvent bath. Doping of the graphene with TFSA was accomplished by spin-casting TFSA (20~mM in nitromethane) at 1000-1500~rpm for 1~min. Ohmic contacts to the Si wafers were made by gallium indium eutectic paint (99.99\% metal basis), and $J$-$V$ and $C$-$V$ measurements were taken between the graphene (metal electrode) and ohmic contact on Si (semiconductor) as shown in \ref{Diagram}c. 

To measure the external quantum efficiency (EQE), the devices are illuminated by monochromatic light mechanically chopped at 400Hz, and the photocurrent is recorded by a Stanford Research System 830 DSP lock-in amplifier together with a Keithley 428 current amplifier. A Xe-arc lamp is used as the white light source and an Oriel monochromator is adopted to generate such monochromatic light. The monochromatic light intensity is measured using a calibrated Newport 818-UV Si detector. To measure the carrier lifetime, transient photovoltages are induced by 632 nm laser pulses with light intensity attenuated by a series of neutral density filters~\cite{TPV}. The resulting voltage signals are then monitored by an oscilloscope.


\begin{figure}[t]
\includegraphics[angle=0,width=0.75\textwidth]{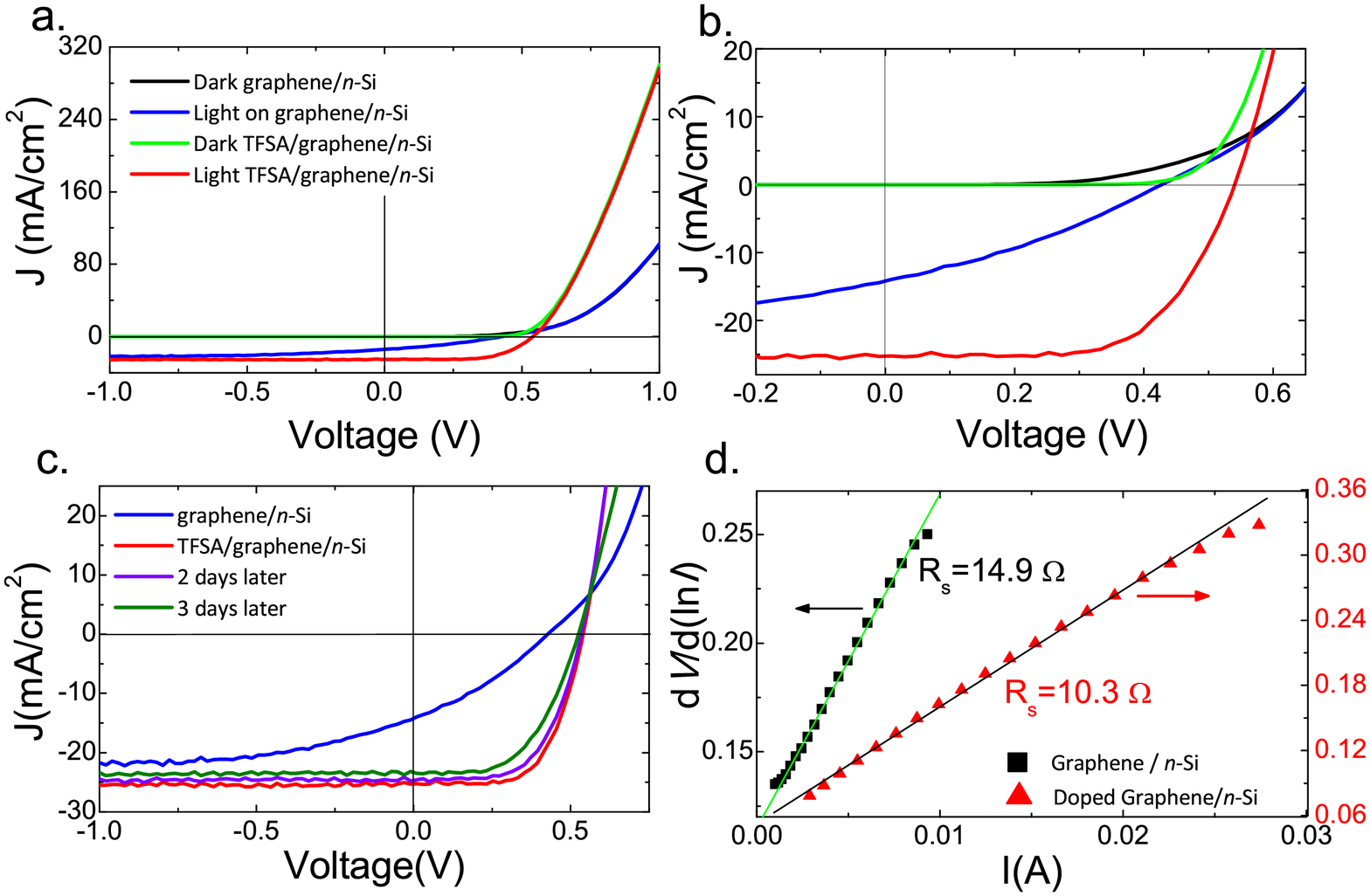}
\caption{a. Current density $J$ versus voltage $V$ curves of graphene/$n$-Si and doped-graphene/$n$-Si Schottky solar cells in dark and after illumination. b. Data of panel a on expanded scales. c. $J$-$V$ plots of graphene/$n$-Si and doped-graphene/$n$-Si junctions under illumination with time. d. The series resistance $R_s$ values extrapolated from $dV$/$d$ln$I$ vs $I$ curves before and after the doping.}
\label{IV}
\end{figure}

The work function difference between the graphene and the \textit{n}-Si results in electrons transferred from the Si to the graphene yielding a Schottky junction with its associated depletion layer in the Si and built-in potential across it. Photons absorbed in the silicon generate electron-hole pairs that are separated within the depletion region by the electric field associated with $V_{bi}$, and the charges are collected at the graphene and semiconductor contacts, thereby generating power from the device. \ref{IV}a-b shows \textit{J}-\textit{V} curves in the dark (black line) and under AM1.5 illumination at 100 mW/cm$^2$ (blue line). Under illumination, the short-circuit current ($J_{sc}$) becomes 14.2~mA/cm$^2$ with open-circuit voltage ($V_{oc}$) and power conversion efficiency (PCE) percentages corresponding to 0.42~V and 1.9\%, respectively. Similar measurements have been taken on 9 different samples with PCE values ranging from 1\% to 2\% and the trends reported here reproduced on all the samples. 

\begin{figure}[h]
\includegraphics[angle=0,width=0.70\textwidth]{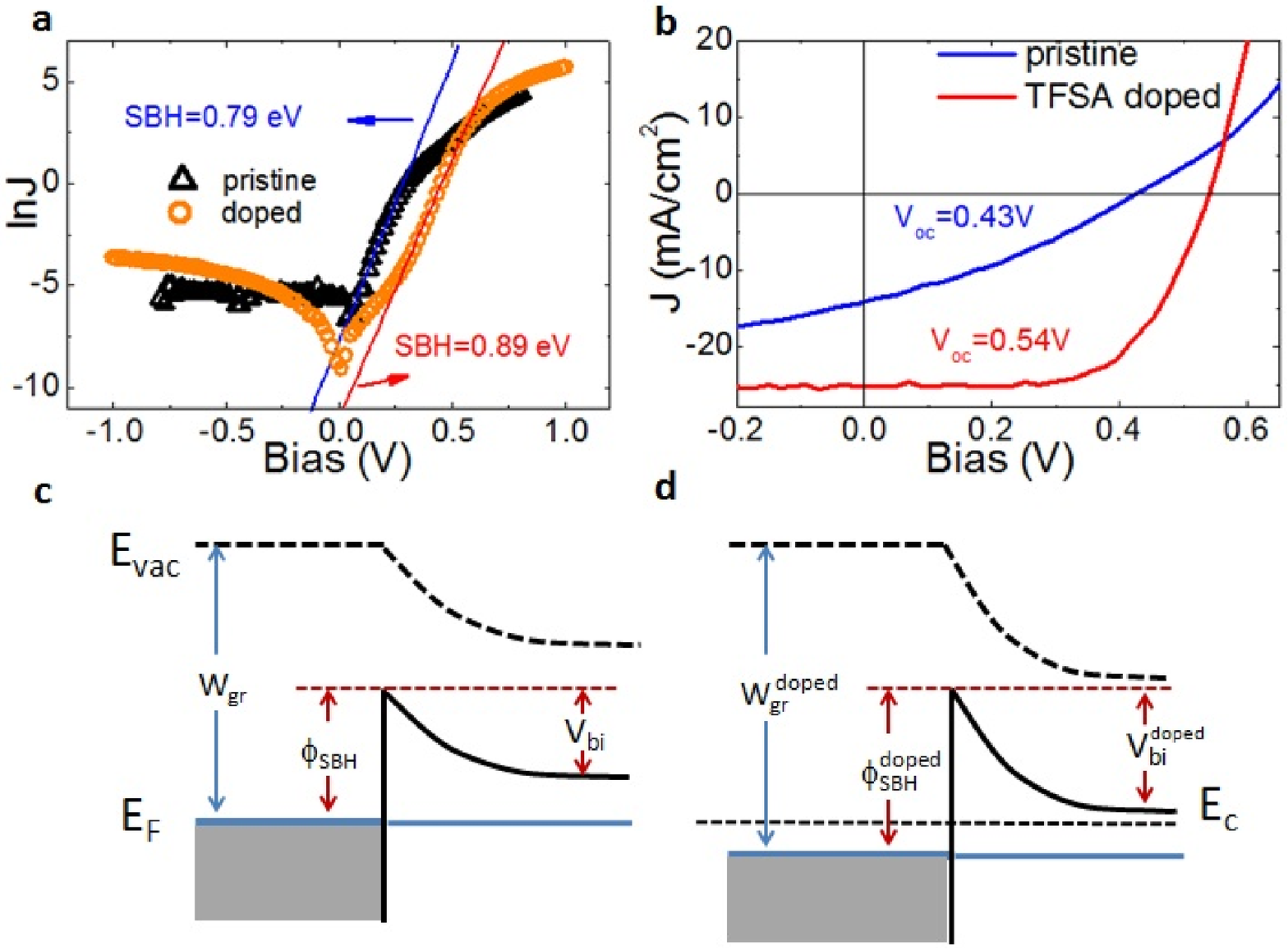}
\caption{a. $J$-$V$ characteristics in the semi-logarithmic scale, b. Zoomed in $J$-$V$ characteristics of graphene/$n$-Si diodes, and c-d the band diagram at the graphene/$n$-Si interface before and after the doping.}
\label{IV2}
\end{figure}

\ref{IV}a-b also shows the $J$-$V$ characteristics after doping the graphene sheets with TFSA. We have shown previously that TFSA hole dopes (\textit{p} dopes) graphene, reducing its sheet resistance while also increasing its work function ($W_{gr}$) without changing its optical properties~\cite{tongaynano}. For the graphene/\textit{n}-Si solar cells upon doping with TFSA, the $J_{sc}$, $V_{oc}$, and fill factor (FF) all increase from 14.2 to 25.3 mA/cm$^2$, 0.43 to 0.54~V, and 0.32 to 0.63, respectively. These increases in $J_{sc}$ and $V_{oc}$ boost the PCE percentage from 1.9\% to 8.6\%, which is the highest PCE reported for graphene based solar cells to date. 

This dramatic improvement can be partially attributed to the improvement in graphene's electrical conductivity~\cite{tongaynano} and the associated reduction of ohmic losses. Plots of $dV$/$d$ln$I$ versus $I$ (\ref{IV}d) allow extraction of the cell series resistance $R_s$.\cite{Cheung,Werner}. From \ref{IV}d, the slopes for pristine and doped graphene/$n$-Si diodes give $R_s$ values of 14.9~$\Omega$ and 10.3~$\Omega$ respectively, a result consistent with TFSA's acceptor nature\cite{tongaynano}. In addition to the reduction of ohmic losses, we note that the increase in $V_{oc}$ implies that the built-in potential $V_{bi}$ also increases, since $V_{oc}$ scales linearly with $V_{bi}$. The built-in potential  $V_{bi}$ is related to $\phi_{SBH}$ via the expression, $\phi_{SBH}=V_{bi}+q^{-1}k_BT$ln$(N_{C}/N_{D})$,~\cite{sze} where $N_C$ is the effective density of states in the conduction band and $N_D$ is the doping level of the semiconductor. Accordingly, an increase in $V_{bi}$ suggests that $\phi_{SBH}$ is increased after doping as shown schematically in panels c and d of \ref{IV2}. According to the Schottky-Mott model, the SBH at the graphene/$n$-Si interface can be related to the difference between the graphene work function W$_{gr}$ and the electron affinity $\chi_{Si}$ of the semiconductor by the equation, $\phi_{SBH}$=$W_{gr}$-$\chi_{Si}$. Since the TFSA hole dopes the graphene electrodes, higher W$_{gr}$ increases the SBH and the greater difference between W$_{gr}$ and $\chi_{Si}$ results in larger charge transfer across the M-S interface, creating a larger potential drop $V_{bi}$ across the depletion width and allowing a more efficient collection of electrons and holes. 

While the increase in $V_{oc}$ provides indirect evidence for an increased SBH and $V_{bi}$, the increase in the SBH by doping is further confirmed by analyzing the ln$J$-$V$ data taken from pristine and doped graphene/$n$-Si solar cells in the dark room environment (\ref{IV2}a). In the ln$J$-$V$ plot, the current density displays adequate linearity over a range of three decades of $J$, and the extrapolation to zero bias yields the saturation current density $J_{s}$ which can be related to the SBH using the thermionic-emission based diode equation,
\begin{equation}  \label{richard}
J(T,V) = J_{s}(T) [\exp ({eV}/{\eta k_B T})-1]   ,
\end{equation}
where $J(T,V)$ is the current density across the graphene/semiconductor interface, \textit{V} the applied voltage, \textit{T} the temperature and $\eta$ the ideality factor~\cite{sze}. The prefactor, $J_s (T)$ is the saturation current density as expressed by $J_{s} = A^* T^2 \exp ({-e \phi_{SBH}}/{k_B T})$, where $\phi_{SBH}$ is the zero bias Schottky barrier height (SBH) and $A^{*}$ is the Richardson constant. Using the $J_s$ values before and after the doping, $J_s^{undoped} = 6.77 \pm 1.74 \times 10^{-4}$ and $J_s^{doped} = 1.05 \pm 0.23 \times 10^{-5}$ mA/cm$^{2}$, in ~\ref{richard}, the Schottky barrier height at the pristine graphene/$n$-Si interface ($\phi_{SBH}^{undoped}$) increases by 0.1~V from 0.79 to 0.89~V after doping ($\phi_{SBH}^{doped}$) consistent with the 0.11~V increase in $V_{oc}$ (\ref{IV}b) and the hole doping nature of TFSA~\cite{tongaynano}. 

While the TFSA doping of graphene/Si junctions gives rise to a 0.1~eV change in $V_{oc}$, we might have expected from our previous results on TFSA doping of graphene the Fermi level shifts to be as high as 0.7~eV~\cite{tongaynano}. We attribute this discrepancy to the fundamental differences between the two experimental setups. In the 4-terminal contact geometry, where the graphene has been transferred directly onto an oxidized Si substrate thereby limiting the charge transfer between the substrate and the graphene, the effect of TFSA doping is optimized. In addition, since current preferentially flows through the most conductive patches (charge puddles exist on the graphene or doped graphene), this measurement therefore overestimates the change in $E_F$. For the graphene/Si junctions reported here, the graphene interacts with the semiconducting substrate and the doping effect is reduced as a result of electronic equilibrium across the interface. By preferentially selecting out regions dominated by low SBHs (i.e., puddles with higher $E_F$) the $J$-$V$ measurements underestimate the change in the Fermi level.

Here, we also note that the undoped graphene/$n$-Si diodes typically yield ideality constants having values in the 1.6 - 2.0 range, which upon doping are improved to values in the 1.3 - 1.5 range. Typically, ideality constants greater than the unity imply a number of possibilities: (1) additional charge transport processes such as thermionic field emission exist at the interface~\cite{tung}, (2) the SBH is bias dependent since the graphene Fermi level is bias dependent~\cite{tongayprx,tongaygan,tung}, (3) the image force lowering~\cite{sze} effect is significant at the interface, and/or (4) Schottky barrier inhomogeneity is present in the junction area~\cite{guttler}. In accord with these possibilities we suggest that for pristine graphene/$n$-Si, ideality constants greater than unity can be associated with the existence of charge puddles on the graphene that are unintentionally formed during the graphene processing steps and which give rise to associated Schottky barrier inhomogeneities. We anticipate that the controlled doping of graphene by TFSA yields more uniformly doped regions\cite{tongaynano}, thereby reducing the Schottky barrier inhomogeneity and giving rise to lower (improved) ideality constants after the doping process. Moreover, we note that the additional \textit{p}-doping associated with the TFSA places the Fermi level of the graphene further away from the neutrality point into a region where the density of states is higher. A higher density of states reduces the bias dependence of the SBH and therefore improves the ideality. 

The 30\% reduction in series resistance $R_s$ of the doped sample is not by itself sufficient to explain the accompanying factor of 1.8 increase in $J_{sc}$. When the device is under illumination, \ref{richard} is modified as
 
\begin{equation} \label{lightJV}
J(T,V) = J_{s}(T) [\exp ({eV}/{\eta k_B T})-1]-J_{ph}
\end{equation}
where $J_{ph}$ is the photon current density, a term usually regarded as a constant for fixed illumination. We have assumed for the purposes of this calculation that $R_s$ is zero so that there are no internal voltage drops. At zero bias ($V = 0$), we have $J_{sc}= - J_{ph}$, and for zero total current ($J = 0$), we have $J_{ph} \approx J_{s}(T) [\exp ({eV_{oc}}/{\eta k_B T})$ where we have assumed that the temperature is high enough to assure that the exponential term dominates. Thus,

\begin{equation} \label{Jph}
J_{sc} \approx -J_{s}(T) \exp ({eV_{oc}}/{\eta k_B T}) = - A^{*}T^2 \exp ({-e \phi_{SBH}}/{k_B T})\exp ({eV_{oc}}/{\eta k_B T})
\end{equation}
By substituting the experimental values of $\phi_{SBH}$, $V_{oc}$ and $\eta$ before and after doping, we calculate that the $J_{sc}$ increases by a factor of 1.5, which is close to the factor of 1.8 gleaned from the $J$-$V$ measurements. However, this value is an underestimate, since we do not take account of the 30\% reduction in series resistance $R_s$ that faciliates charge collection nor do we include any beneficial anti-reflection coating effect of the TFSA overlayer.
\begin{figure}
\includegraphics[angle=0, width=0.90\textwidth]{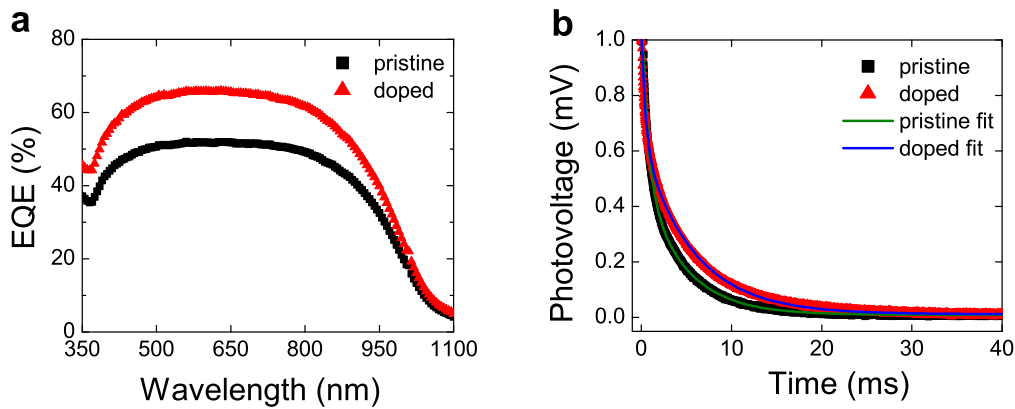}
\caption{(a) External quantum efficiency (EQE) vs wavelength ($\lambda$, nm) and transient photovoltage as a function of time for pristine and TFSA-doped graphene/\textit{n}-Si solar cells.}
\label{EQE}
\end{figure}

\begin{figure}[t]
\includegraphics[angle=0,width=0.60\textwidth]{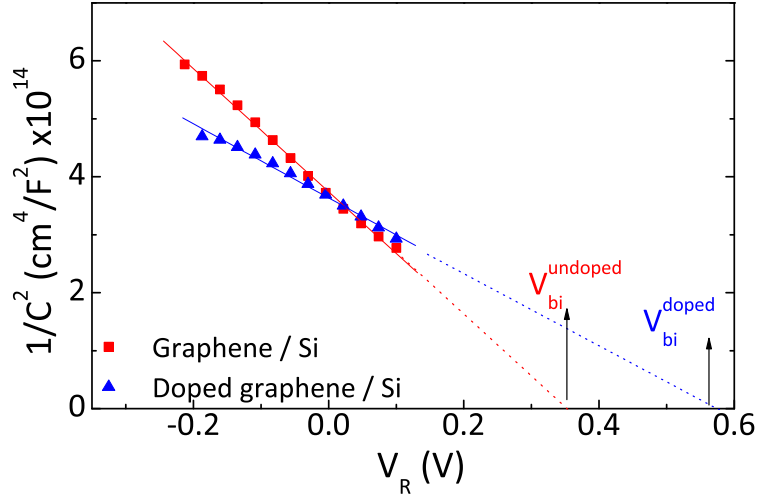}
\caption{Room temperature inverse square of the capacitance (1/$C^{2}$) versus reverse bias ($V_R$) plots before and after the TFSA doping.}
\label{CV}
\end{figure}

Additional characterization is presented in ~\ref{EQE}(a) which shows the external quantum efficiency (EQE) before and after doping in a separate device with PCE increasing from 2.7\% to 6.2\%. The EQE of the pristine cell is similar to state-of-the-art Si solar cells~\cite{SiEQE} because in our sample with its transparent graphene electrodes only the Si absorbs photons which create the electron-hole pairs. The pristine device shows an EQE near 50\% for wavelengths in the range 400~nm < $\lambda$ < 850~nm, indicating significant electron-hole
pair generation and the subsequent facile collection of electrons and holes by the corresponding
electrodes. After TFSA-doping, the EQE was significantly increased to values over 60\% within the above-mentioned photo-responsive range, representing a $\sim$30\% enhancement compared to the pristine cell. Since the photo-generation for the device before and after doping is identical, the higher EQE for the doped device is due to more efficient charge separation and charge collection as a result of increased SBH and reduced R$_s$.

We have also employed (~\ref{EQE}(a)) a transient photovoltage technique to study the dissociated charge carrier lifetimes. By fitting the photovoltage response with two exponential decay components for Si solar cells as done in previous work~\cite{Zhao1996, Cui2011}, we can extract out the carrier lifetime both at graphene/Si interface ($\tau_1$) and inside the bulk Si ($\tau_2$). As shown in ~\ref{EQE}(b), the experimental data are well described by ~\ref{lifetime}. 
\begin{equation}\label{lifetime}
V_{photo}=V_1\textrm{exp}(-\tau_1/t)+V_2\textrm{exp}(-\tau_2/t)+V_0  ,
\end{equation}
where $V_1$ and $V_2$ are the magnitudes of the two exponential components and $V_0$ is a background term.
For our pristine device, $\tau_1^{undoped}$=0.71 $\pm$ 0.01~ms and $\tau_2^{undoped}$=4.49 $\pm$ 0.02~ms. Such long carrier lifetimes are expected since we are using crystalline Si with a high carrier mobility. After doping, $\tau_1^{doped}$=0.46 $\pm$ 0.01~ms and $\tau_2^{doped}$=5.68 $\pm$ 0.01~ms. Here $\tau_1$ is probably related to the interface property between graphene and Si that is not altered significantly by TFSA, and $\tau_2$ is determined by the bulk property of Si which unlikely to be affected by TFSA. Therefore, carrier lifetime should not be changed remarkably neither at interface nor inside the bulk as observed experimentally. In addition, such long carrier lifetime also indicates it not a main factor to the significant improvement of the device performance. Hence, the enhancement in the photocurrent and thus the overall PCE is mainly a consequence of the increase in the SBH and the decrease in R$_s$ as manifested in the noticeable increase in EQE.

While $J$-$V$ characteristics allow us to estimate the change in the SBH, analysis of the capacitance versus bias ($C$-$V$) measurements provides information about the magnitude of $V_{bi}$. Within the Schottky Mott relationship, the diodes in reverse bias satisfy the relation, $1/C^2=2(V_R+V_{bi})/eN_D\epsilon_{s}\epsilon_0$. In this model, $C^{-2}$ scales linearly with V$_R$ (~\ref{CV}) and extrapolation to the abscissa yields $V_{bi}$. From ~\ref{CV}, we observe that $V_{bi}$ increases by 0.2~V, from 0.36 to 0.56~V after doping. The $V_{bi}$ values extracted from the $C$-$V$ measurements yield SBHs of 0.63~V before and 0.82~V after the TFSA doping in accord with the SBH values extracted from the $J$-$V$ characteristics for a different sample (\ref{IV2}) and the observed increase in $V_{oc}$ (\ref{IV}). We note that these two measurements, $J$-$V$ and $C$-$V$, provide complementary techniques for determining the SBH and $V_{bi}$. While $J$-$V$ measurements manifest current transport processes across the graphene/$n$-Si, capacitance measurements probe the space-charge region of the Schottky junction. As a result of these fundamental differences and in the presence of Schottky barrier inhomogeneities\cite{guttler}, $J$-$V$ characterization measures the lowest SBH while $C$-$V$ measurements provide an average SBH at the interface\cite{sze,tung} resulting in different $V_{bi}$ values. 

The environmental stability of the solar cells described here relates to our previous observation that, owing to the hydrophbic nature of TFSA, TFSA-doped graphene displays superior stability while preserving graphene's optical properties.~\cite{tongaynano}. However, in ambient atmosphere various gases such as O$_2$ intercalate beneath the graphene sheets transferred onto various materials~\cite{sutter}. Particularly, O$_2$ intercalation between the graphene/Si interface can be problematic since further oxidation of the Si surface results in tunnel barriers that increase ohmic losses within the sample, leading to a gradual decrease in the overall power conversion efficiency during a week. We find that spin-casting a PMMA polymer layer on graphene/$n$-Si and TFSA-doped graphene/$n$-Si solar cells not only inhibits this additional oxidation (over a typical two-week observation time) but also slightly improves the PCEs, since the PMMA, possibly in combination with TFSA, acts as an anti-reflection coating. However, $J$-$V$ measurements taken on graphene/$n$-Si and PMMA coated graphene/$n$-Si show no change in the $J$-$V$ characteristics, implying that the PMMA itself does not dope the graphene electrodes.

To summarize, we have shown improved light harvesting in chemically doped graphene/$n$-Si Schottky junction solar devices. Doping with TFSA overlayers results in an $\sim$4-5 times increase in power conversion efficiencies of the graphene/$n$-Si Schottky junction solar cell junctions from 1.9\% to 8.6\%. We attribute the improvement in the PCE to reduction of graphene's sheet resistance and hence of $R_s$ and to an increase in the built-in potential $V_{bi}$ as measured by three complementary techniques, $J$-$V$, $C$-$V$ and EQE measurements. While the improvement in the graphene sheet resistance reduces the ohmic losses within the sample, an increase in $V_{bi}$ more efficiently separates the electron-hole pairs generated by absorbed photons. The methods described to accomplish this are practical, simple and scalable since device fabrication involves simple planar thin-film geometries, conventional graphene production techniques and uncomplicated spin-casting of organic layers. 

\acknowledgement

The authors greatly appreciate significant technical support and useful discussion with Renjia Zhou and Dr. Jiangeng Xue from Department of Material Science and Engineering at University of Florida. This work is supported by the Office of Naval Research (ONR) under contract number 00075094 (BRA) and by the National Science Foundation (NSF) under Contract Number 1005301 (AFH).
\suppinfo

\bibliography{acs_submission}

\end{document}